\documentstyle[editedvolume,epsf]{crckapb} 
 \def\mso{\, M_\odot}
 \def\rso{\, R_\odot}
 \def\lso{\, L_\odot}
 
 \def\kms{\, {\rm km}\, {\rm s}^{-1}}
 \def\teff{\log\, T_{\rm eff}\,}
 
 \def\ra{\rightarrow}
 \def\simle{\mathrel{\hbox{\rlap{\hbox{\lower4pt\hbox{$\sim$}}}\hbox{$<$}}}}
 \def\simgr{\mathrel{\hbox{\rlap{\hbox{\lower4pt\hbox{$\sim$}}}\hbox{$>$}}}}
 \def\msoy{\, \mso~{\rm yr}^{-1}}

\def\vrot{v_{\rm rot}}
\def\vcrit{v_{\rm crit}}

\def\teff{T_{\rm eff}}
\newcommand{\oc}{{\omega_{\mathrm{c}}}}
\newcommand{\ok}{{\omega_{\mathrm{Kep}}}}

\begin{opening}
\title{B[e] SUPERGIANTS: WHAT IS THEIR EVOLUTIONARY STATUS?}
\subtitle{}
\author{N. LANGER}
\institute{Institut f\"ur Physik, Universit\"at Potsdam\\
           Postfach 601553, D--14415 Potsdam, Germany}
\author{A. HEGER}
\institute{Max-Planck-Institut f\"ur Astrophysik\\
          D--85740 Garching, Germany}
\end{opening}
\runningtitle{THE EVOLUTIONARY STATUS OF B[e] STARS}
\begin{document}
\begin{abstract}
In this paper, we investigate the evolutionary status of B[e]~stars
from the point of view of stellar evolution theory. We try  
to answer to the question of how massive hot supergiants
--- i.e. evolved stars --- can be capable of producing a
circumstellar disk. We find and discuss three possibilities: 
very massive evolved main sequence stars 
close to critical rotation 
due to their proximity to their Eddington-limit,
blue supergiants which have just left the red supergiant branch,
and single star merger remnants of a close binary system. While the latter
process seems to be required to understand the properties of the
spectroscopic binary R4 in the LMC, the other two scenarios 
may be capable of explaining the distribution of the B[e] stars
in the HR~diagram. 
The three scenarios make different
predictions about the duration of the B[e]~phase, the
time integrated disk mass and the stellar properties during the B[e]~phase,
which may ultimately allow to distinguish them observationally.
\end{abstract}

\section{Introduction}

In recent years, the so called B[e] supergiants (cf. Lamers, this volume)
have emerged as a distinct class of massive stars, whose 
defining properties --- a strong mid-IR excess, strong Balmer emission,
and narrow permitted and forbidden low-excitation emission lines ---
can be explained by a slowly outflowing equatorial disk superimposed to a 
normal fast wind (cf. Zickgraf et al. 1996a).
Although the number of these objects is small --- the best statistics
exists for the Magellanic Clouds, where about 15 B[e] supergiants
have been found (Gummersbach et al. 1995) --- it is comparable
to the number of Luminous Blue Variables (LBVs; cf. Bohannan 1997),
which are know to represent a key evolutionary phase of very massive
stars (Langer et al. 1994, Garc\'{\i}a-Segura et al. 1997,
Langer et al. 1998). It is thus of 
fundamental importance to understand whether the B[e] supergiants
are just freaks, i.e. peculiar objects which come to exist
due to special circumstances --- in which case they might still
show interesting physical phenomena --- or whether all stars
within a certain initial mass range evolve through a B[e] phase.
In this case, our general understanding of the evolution of stars
in that mass range may depend on our understanding of the B[e] stars.

Although we have a solid knowledge of the evolution of
the deep interior of massive
stars since a long time (Weaver et al. 1978, Kippenhahn \& Weigert, 1990)
it has become
more and more clear during the last decades that our understanding of the
evolution of their observable features is still rather incomplete.
Among others, open problems concern the effective temperature evolution
of moderately massive ($M_{\rm ZAMS} \simeq 10 ... 30\mso$)
post main sequence stars
(Langer \& Maeder 1995) and the evolutionary connections between O~stars,
Luminous Blue Variables (LBVs) and Wolf-Rayet (WR) stars
for higher masses (Schaller et al.
1992, Langer et al. 1994, Stothers \& Chin 1996, Pasquali et al. 1997;
see also Schulte-Ladbeck 1998).
Two major physical difficulties in the theoretical models of the
observable evolutionary stages of massive stars have been identified
and made responsible for the persisting lack of reliable models:
mass loss and internal mixing processes (Meynet et al. 1994, Langer 1994,
Deng et al. 1996). E.g.,
it has been found that the mass loss of massive main sequence stars
should be roughly twice as high as what appears to be observed in order
to understand many features of massive post main sequence stars (Meynet
et al. 1994, Langer et al. 1994).
Additionally, there is growing evidence that stellar
rotation may considerably affect the evolution of massive stars
(Maeder 1987, Langer 1991a, Fliegner et al. 1996, Maeder \& Meynet 1996,
Meynet \& Maeder 1997, Langer et al. 1997ab).
{\it Rapid rotation} can reduce
the effective gravity in the star, and it produces large scale flows
(Eddington 1925). During the evolution, {\it differential rotation}
occurs in all stars, with the possibility of the occurrence of various local
hydrodynamic instabilities (cf. Endal \& Sofia 1978, Zahn 1983)
and corresponding mixing of chemical elements and angular momentum.
Of relevance for massive stars are the shear instability (cf. Maeder 1997),
the baroclinic instability (Zahn 1983, Spruit \& Knobloch 1984),
and the Solberg-H{\o}iland and Goldreich-Schubert-Fricke instabilities
(cf. Korycansky 1991).

Time dependent evolutionary models for massive stars including rotation
have been constructed in the past in one dimension, using various degrees
of approximation (e.g. Endal \& Sofia 1978, Maeder 1987, Langer 1991a,
Langer 1992, Talon et al. 1997, Langer 1998; cf. also Dupree 1995).
Today, it is beyond reasonable doubts that the evolution of massive stars
is influenced by rotation due to the physical mechanisms mentioned
above (cf. Fliegner et al. 1996). While the principle effects of
rotation in the interior of massive stars during their evolution all the
way to iron core collapse are described elsewhere (Langer et al. 1997b,
Heger et al. 1998), we discuss here whether
massive single stars can approach the limit of critical rotation.
We shall assume in Sections~2 and~3 that this would lead to the disk
structure responsible for the B[e] phenomenon, either due to the
Bjorkman-Cassinelli (1993) mechanism of wind compression of a
rotating star or otherwise. A detailed wind disk model for B[e] stars can be
found in Bjorkman (this volume), while the recent status of the
Bjorkman-Cassinelli model is discussed by Owocki et al. (1996),  
Owocki (this volume) and Cassinelli (this volume).
In Section~4 we will discuss a model where the disk
is not produced by a critically rotating single star
but rather a critically rotating close binary.

\section{Very massive main sequence stars}

Massive main sequence stars are rapid rotators, with equatorial rotation
velocities in the range of $100 ... 400\kms$ (Fukuda 1982, Penny 1996,
Howarth et al. 1997).
In our first scenario, we investigate whether massive main
sequence stars, i.e. massive stars during core hydrogen
burning, are capable of arriving a critical rotation (cf. Langer 1998).
The results of this Section have been obtained with
a hydrodynamic stellar evolution code (cf. Langer et al. 1988, Langer 1991b),
employing the OPAL opacities of Iglesias et al. (1992). We have used an
outer boundary condition which takes the optical depth of the stellar
wind selfconsistently into account within a grey approximation
(Langer et al. 1994, Heger \& Langer 1996).

%=============== Figure v200 =========================================
\begin{figure}[t]
\centering
\epsfxsize=0.9\hsize
%\mbox
\centerline{\epsffile{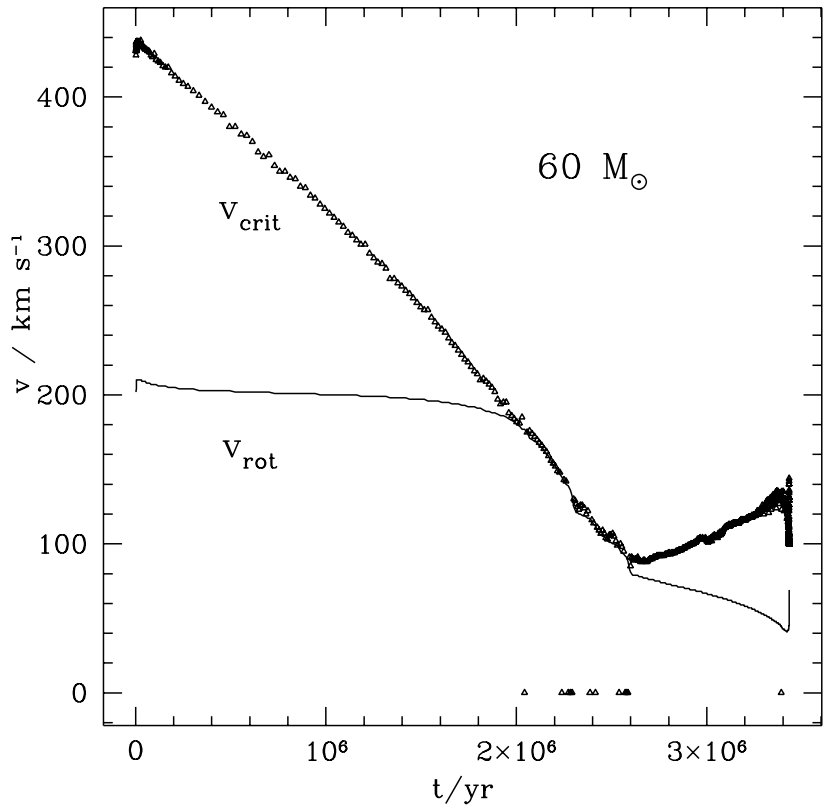}}
\caption[]{Equatorial rotational velocity $\vrot$
 and the critical rotational velocity $\vcrit$
 as function of time during the main sequence evolution of
 a $60\mso$ sequence.
}\end{figure}
%===================================================================

We have computed the proximity of the star to the Eddington-limit,
i.e. the Eddington factor $\Gamma = L/L_{\rm edd} =
\kappa L / (4 \pi c G M)$ in the following way. It has been shown
in Langer (1997) that the occurrence of convection and of density inversions
makes the concept of the Eddington limit as a stability limit invalid
in the stellar interior, i.e. the Eddington factor $\Gamma$ has to be
evaluated only at the stellar surface. Since the term ``surface'' is
not unambiguously defined in this context, we considered
$\Gamma$ in layers with an optical depth of $\tau < 100$, where in fact
neither a significant
convective energy flux nor density inversions have been found in
the investigated models. Thus, to estimate the distance to the Eddington
limit, we used the maximum value of $\Gamma$ occurring in the subsonic
layers with $\tau < 100$. Furthermore, we used the OPAL opacity
coefficient to compute the Eddington factor. Mass and luminosity
are practically constant for  $\tau < 100$ and equal to the total
stellar mass and luminosity.

%=============== Figure m200 =========================================
\begin{figure}[t]
\centering
\epsfxsize=0.9\hsize
%\mbox
\centerline{\epsffile{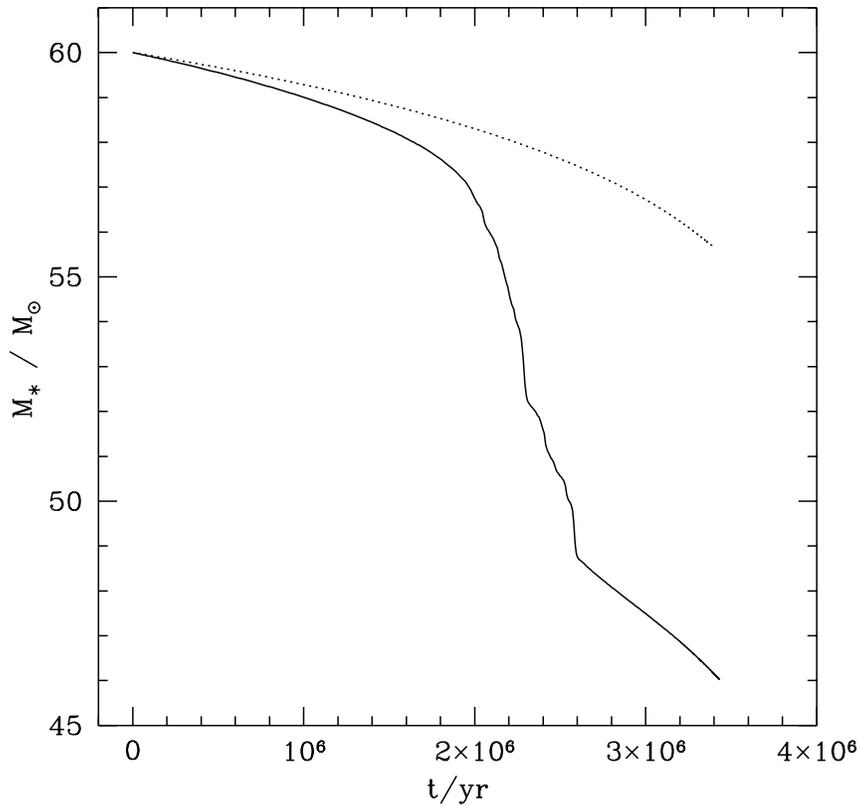}}
\caption[]{Stellar mass as a function of time for the 60$\mso$
  sequence shown in Fig.~1 (solid line),
  in comparison to the evolution of a 60$\mso$ star without
  enhancement of the mass loss rate due to rotation (dotted line).
  The mass loss rates can be read of this figure as the
  slope of the curves.}
\end{figure}
%===================================================================

Angular momentum is carried only as a passive
quantity in the stellar models, 
i.e. we ignore the centrifugal force in the stellar interior
as well as the effect of mixing of chemical species due to rotationally
induced instabilities (cf. Fliegner et al. 1996,
Meynet \& Maeder 1997). However, we do consider
the centrifugal force at the stellar surface to evaluate the distance
of the star from the $\Omega$-limit (Langer 1997), i.e. from
critical rotation, with
$\Omega = \vrot / \vcrit$,
and
$v_{\rm crit}^2 = G M (1-\Gamma )/R$.
Furthermore, we assume our models to be always rigidly rotating.
Stellar models including differential rotation (e.g. Fliegner et al. 1996)
show that this is a good approximation on the main sequence, since the
time scale for angular momentum transport is of the order of the
thermal time scale and also shorter than the
time scale of rotationally induced chemical mixing
(Chaboyer \& Zahn 1992, Zahn 1992, Talon \& Zahn 1997).
According to Zahn (1994), the expected amount of differential rotation
in a massive main sequence star is roughly
$\Delta \omega / \omega \simeq \omega^2 R^3 / (G M)$,
with $\omega$ being the mean angular velocity and
$\Delta \omega $ its difference between stellar core and surface.
In the models presented below,
this estimate gives $\Delta \omega / \omega \simle 0.01$.
The approximation of rigid rotation
may become invalid when the mass loss time scale
becomes shorter than that of angular momentum transport, but this
is not the case in our models.

%=============== Figure Omega ========================================
\begin{figure}[t]
\vskip 0.7cm
\centering
\epsfxsize=0.9\hsize
%\mbox
\centerline{\epsffile{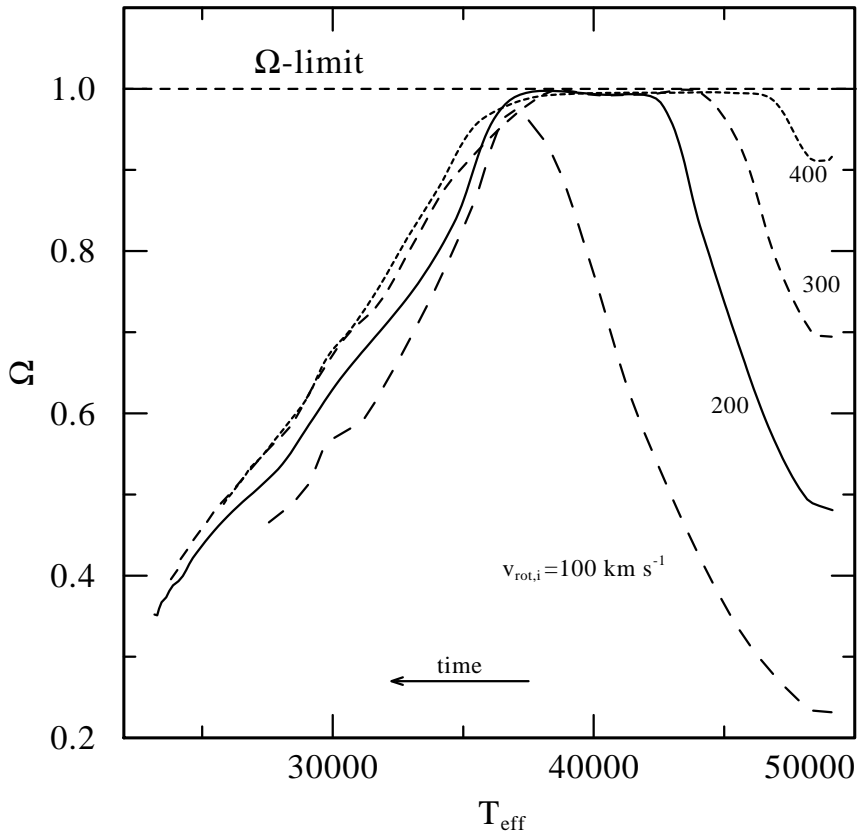}}
\caption[]{Evolution of the ratio of the rotation rate to
the critical rotation rate,
$\Omega$, as function of the effective temperature
during core hydrogen burning, for 60$\mso$ sequences
starting with different equatorial rotational velocities $v_{\rm rot,i}$.
}\end{figure}
%===================================================================

Angular momentum loss is only considered through the effect of mass loss,
i.e. the lost mass carries away its specific angular momentum.
To compute the mass loss rate for our stellar models, we have applied
the empirical rate found by Lamers \& Leitherer (1993) with the
metallicity dependence obtained by Leitherer \& Langer (1991).
However, we have applied the correction factor derived
by Bjorkman \& Cassinelli (1993) as a fit to the results of
Friend \& Abbott (1986) to take the effect of rotation on
the mass loss rate of hot star winds into account
(cf. Langer 1998).

Fig.~1 shows the time evolution of the critical rotational velocity
and the actual rotational velocity for a 60$\mso$ sequence
starting with $\vrot = 200\kms$.
The critical rotational velocity has a pronounced minimum
at roughly $t=2.6\, 10^6\,$yr, which corresponds to a stellar effective
temperature of $\teff\simeq 36\, 500\,$K, around which the iron opacity
peak has its maximum effect. We see in Fig.~1 that, as $\vrot$
approaches $\vcrit$ (i.e. $\Omega\rightarrow 1$) the rotation rate of
the star declines such that the $\Omega$-limit $\Omega =1$ is never
exceeded. 

The reason is that, using the results of Friend \& Abbott (1986),
the mass loss rate increases as the star approaches $\Omega =1$.
In Section~3, we discuss that the angular momentum loss rate is
directly coupled to the mass loss rate of the star (cf. Langer 1998,
and Fig.~6 below). Consequently, the mass loss rate at the 
$\Omega$-limit is determined by the angular momentum loss rate which
is required to prevent the star from reaching $\Omega =1$.

The result is that the star evolves along the $\Omega$-limit until
the critical rotation rate increases again, due to changes in the
photospheric parameters (i.e., the opacity). For the considered
example, the time spent at the $\Omega$-limit is roughly $6\, 10^5\,$yr.
Fig.~2 shows that during this time the mass loss rate is increased to
roughly $10^{-5}\msoy$, which is a factor of $\sim 10$ above the
normal radiation driven mass loss rate (cf. Fig.~2).

According to this picture, very massive main sequence stars may thus,
due to their proximity to their Eddington-limit, evolve at critical
rotation for several $10^5\,$yr. Fig.~3 shows that, for a given star,
this time scale depends on its initial rotation rate. As during the
time at the $\Omega$-limit the star may have a slow dense equatorial
outflow, a relationship to the B[e] stars may be suggested.

\section{Supergiants on a blue loop}

In this Section we describe a possibility for a star to arrive at
critical rotation which does not require an extremely high 
stellar luminosity. It was found to occur in contracting stars,
whose envelope structure changes from convective to radiative
(see Heger \& Langer 1998). This situation occurs for massive stars
undergoing a so called blue loop, i.e. core helium burning
red supergiants which evolve off the Hayashi-line towards the
regime of blue supergiants in the HR diagram. Blue loops are
typically found in the initial mass range $5...25\mso$ (cf.
Langer 1991b, Schaller et al. 1992). However, the mechanism
discussed in the following may also apply to other situations,
e.g. to the post-AGB phase of low mass stars.

Fig.~4 shows evolutionary tracks for rotating stars between
10 and 20$\mso$ (cf. Heger et al. 1997 for details), 
two of which evolve through a blue loop. The 12$\mso$ sequence
will be investigated in more detail below. Figure~5 shows the
time evolution of the radii of various Lagrangian mass shells
and of the stellar radius, before and during the blue loop.
Also shown is the radial extent of the convective part of the
envelope. We see from Fig.~5 that, within a fair approximation,
the inner and the outer radius of the convective
envelope remain constant until the blue loop occurs, while
mass shells continuously flow out of the convection zone through
its lower boundary.

%=============== Figure HRD ========================================
\begin{figure}
\centering
\epsfxsize=0.9\hsize
\centerline{\epsffile{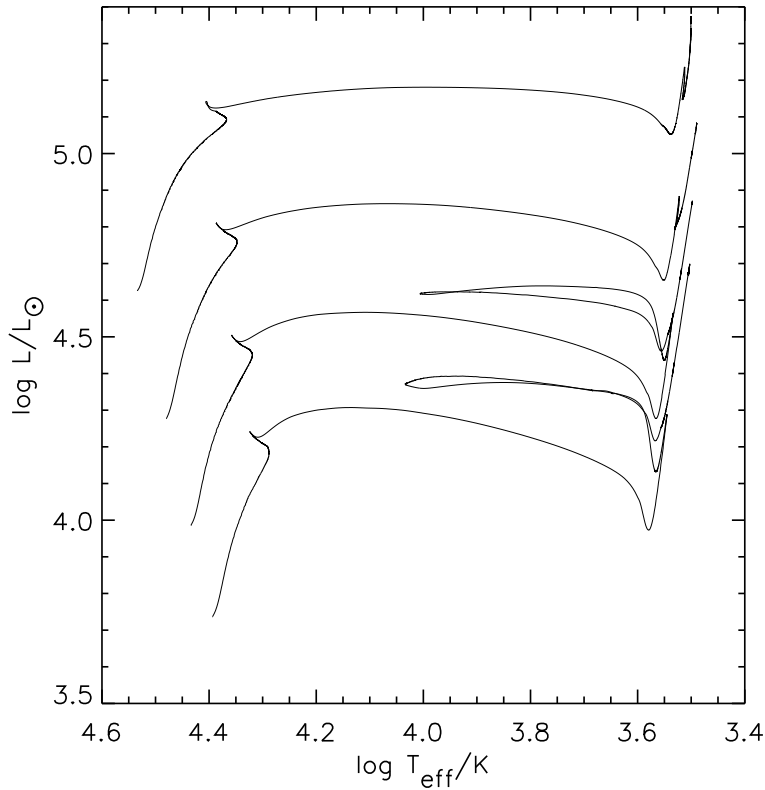}}
\caption[]{Evolutionary tracks for rotating
  $10$, $12$, $15$ and $20\,\mso$
  sequences from the ZAMS to central neon exhaustion
  (cf. Heger et al. 1997). }
\end{figure}
%===================================================================

%=============== Figure ENVELOPE========================================
\begin{figure}
\centering
\epsfxsize=0.9\hsize
\centerline{\epsffile{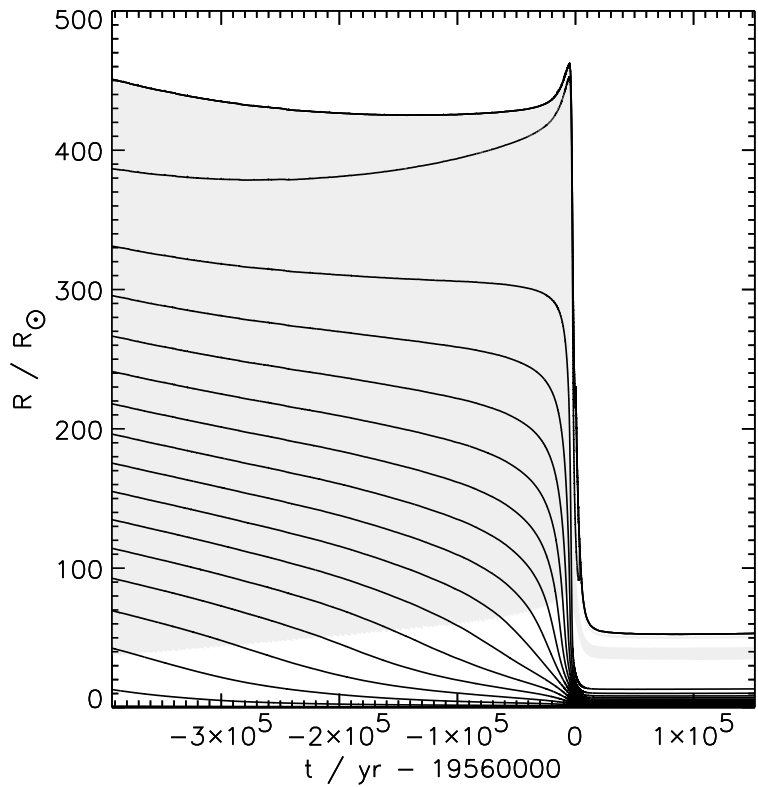}}
\caption[]{Evolution of the radii of different mass shells
  as a function of time for a period including
  the transition from the red to the blue supergiant stage
  of our $12\,\mso$ model. $t=0$ is defined as in Fig.~7
  and corresponds roughly to the time of the red-blue transition.
  Except for the uppermost solid
  line, which corresponds to the surface of the star, the lines trace
  Lagrangian mass coordinates.
  The mass difference between the lines is $0.5\,\mso$.
  Shading indicates convective regions.
  }\end{figure}
%===================================================================

In Fig.~6 we show that, due to the rapid angular momentum transport
within convection zones, this situation leads to a dramatic spin-up
of the envelope. In the right part of Fig.~6 (B1...B4), the situation
is sketched for the assumption of rigid rotation in convective regions
and broken up into three discrete steps. For
a constant inner radius of the convection zone, the drop-out of mass
through its lower boundary leads to an increase of the rotation
frequency and of the mean specific angular momentum in the convection
zone. Note that rigid rotation is assumed only for simplicity; it is
not necessary for the spin-up mechanism to work.

The left side of Fig.~6 (A1...A4) sketches the situation of the loss
of mass through the upper boundary of a rigidly rotating region.
In this case, which applies to the mass loss of very massive main
sequence stars (cf. Sect.~2), matter with the highest specific
angular momentum is continously lost, which leads to a spin-down
of the rigidly rotating star. 

%=============== Figure Scheme========================================
\begin{figure}
\centering
\epsfxsize=0.7\hsize
\centerline{\epsffile{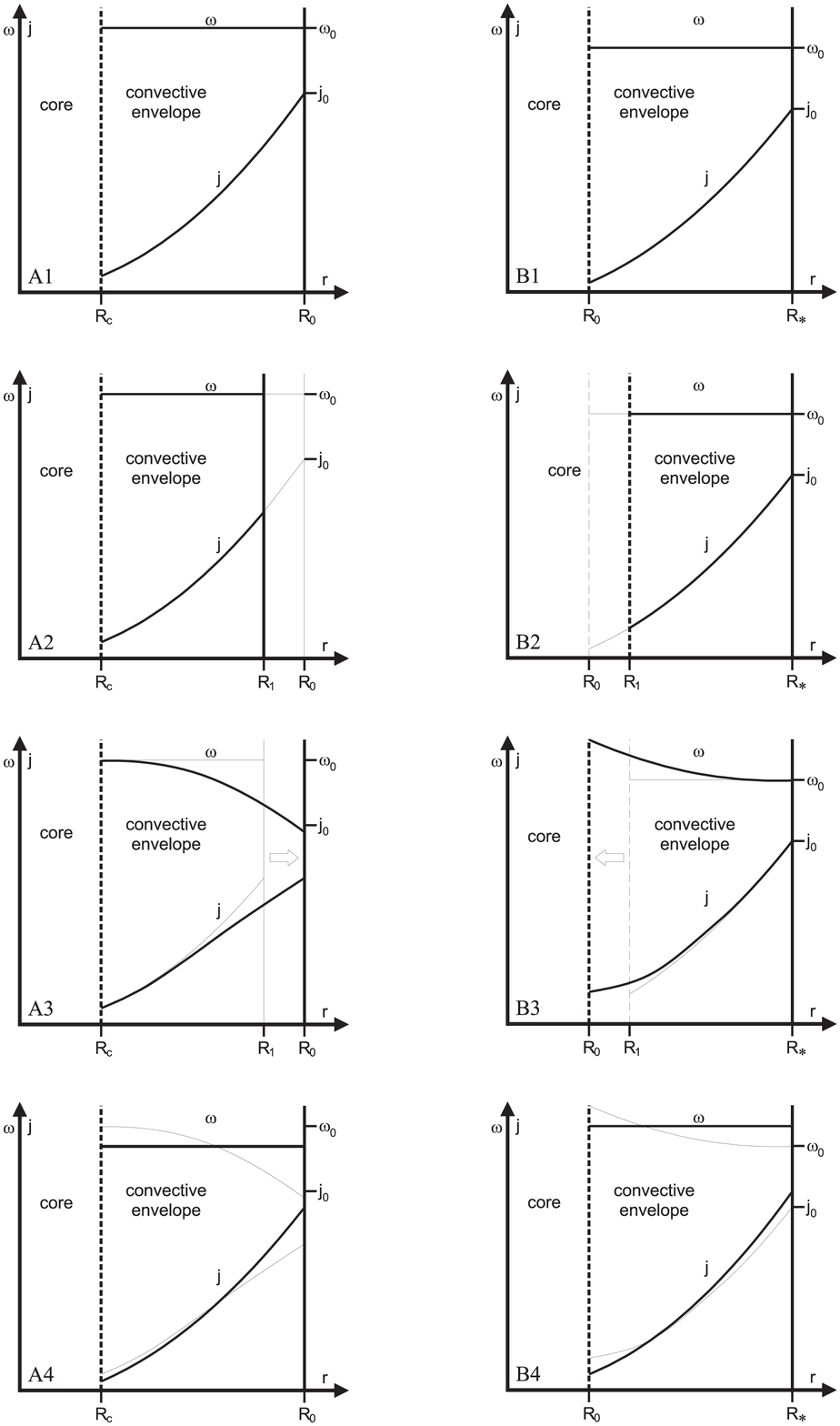}}
  \caption[]{Mass loss from a rigidly rotating stellar envelope from
  the surface (case A: left panels) and through its lower boundary
  (case B: right panels).  The continuous process is 
  split up into three steps. First (panels $1 \to 2$), mass gets lost from
  the envelope, secondly, the envelope restores its original 
  (inner or outer) radius $R_{\rm 0}$
  (panels $2 \to 3$) by expansion, and third (panels $3 \to 4$) the
  specific
  angular momentum $j$ is redistributed such that rigid rotation 
  (i.e. $\omega(r)=$const.) is
  restored.  This leads to spin-down (spin-up) and decrease (increase)
  of the mean specific angular momentum for the case of mass loss through
  the upper (lower) boundary of the rigidly rotating stellar
  envelope.  Thin lines show the state of the preceding step.}
\end{figure}
%===================================================================

%=============== Figure vrot ==========================================
\begin{figure}
\centering
\epsfxsize=0.9\hsize
\centerline{\epsffile{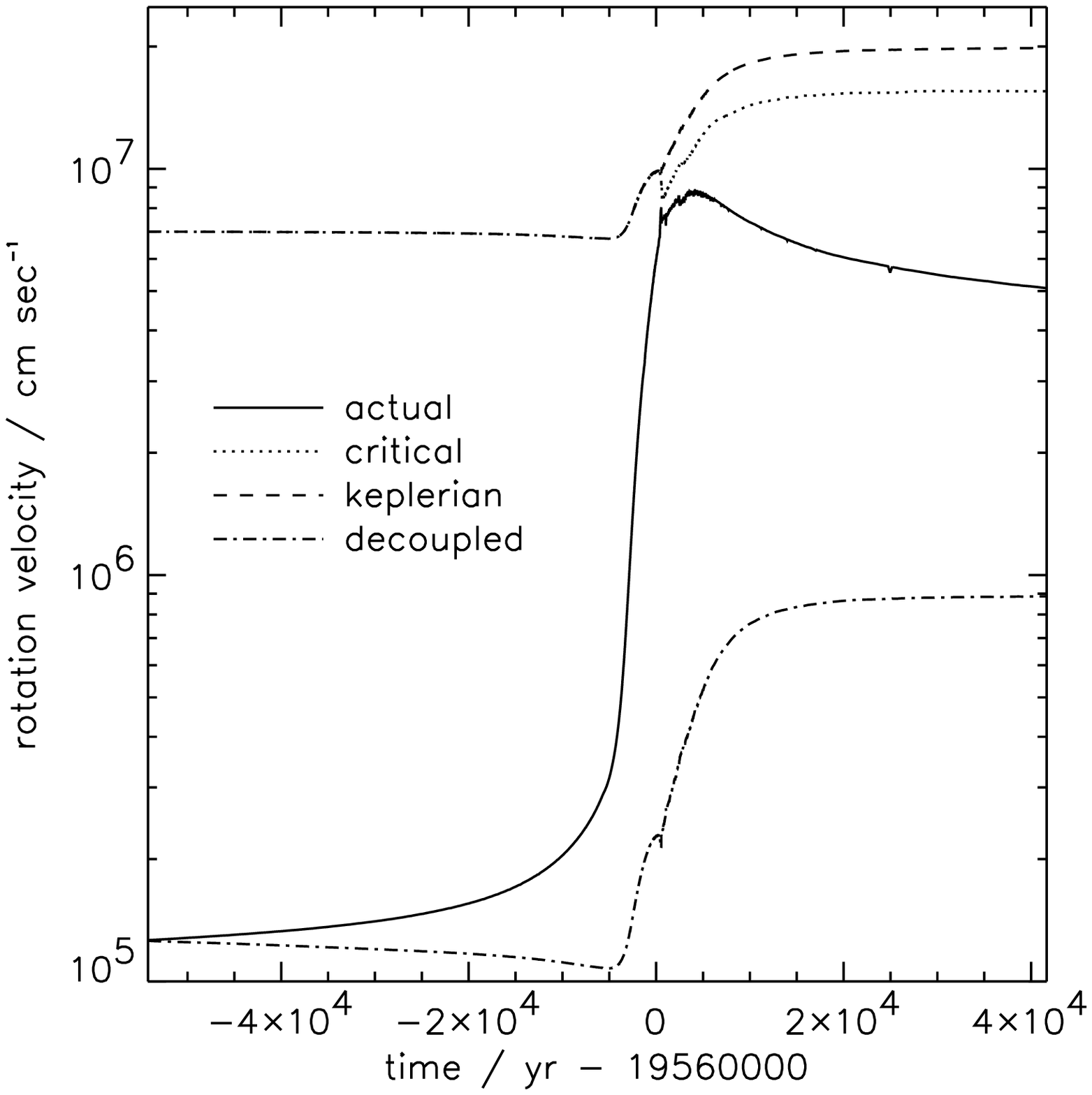}}
  \caption[]{Equatorial rotation velocity as a function of time (solid
  line) compared to the Keplerian (dashed line) and the critical
  (dotted line) rotation rate; the latter two are different by the
  factor $1-\Gamma$.  During the red supergiant phase it is $\Gamma
  \ll 1$ and the two lines coincide, while during the blue supergiant
  phase $\Gamma$ rises to $0.4$.  The dash-dotted line shows the
  evolution of the surface rotation rate if there were no angular
  momentum transport in the convective envelope.}
\end{figure}
%===================================================================

Fig.~7 shows the time dependence of the rotational velocity of
the 12$\mso$ sequence before, during, and after the red-blue transition.
The transition itself takes only about
$10\,000\,$yr.  During this time the angular momentum transport to
the surface layers of the star continues. 
The envelope layers remain rigidly rotating,
continuing to shovel up angular momentum to the surface.  The
contraction of the star by a factor of $f \approx 10$ would increase
the rotational velocity by the same factor
if $j$ were conserved locally (cf. Fig.~7, ``decoupled''). 
 
Since $1-\Gamma \not\ll 1$, the Keplerian angular velocity
$\ok = \sqrt{GM / R^3} \propto f^{3/2}$ can be used as a rough
approximation for the critical rotation frequency $\oc$.  
Thus, $\Omega$ scales as
$f^{1/2}$, and the mere contraction, without any angular momentum transport,
would bring the star by a factor $\sim 3$ closer to its
critical rotation rate.  The numerical value of $\Gamma \approx
2\,10^{-3}$ on the RSG found in our calculation and $\Gamma \approx
0.4$ on the BSG would lead to an increase of $\Omega$ by $\sim
3.9$ from the RSG to the BSG, again assuming that $j$ were conserved
locally.  Actually, the star gets from $\Omega\approx 0.01$ 
on the red supergiant branch at the 
beginning of central helium burning to critical rotation 
($\Omega = 1$) during the
RSG/BSG transition, even though a considerable loss of angular momentum
due to stellar wind mass loss is included in our model (cf. Fig.~7).

In fact, the model sequence would have exceeded the $\Omega$-limit
if we would not have applied a mass loss increase for $\Omega\ra 1$
as in the massive main sequence models described in Section~2.
I.e., like in that case, we find that stars undergoing a blue loop
are likely to arrive at critical rotation, and they may consequently
develop a slow equatorial outflow, which makes them candidates
for B[e] stars. However, in contrast to the very massive star 
models discussed in Sect.~2, the B[e]-phase according to the
blue loop scenario is much shorter, i.e. only some $10^4\,$yr,
and correspondingly the amount of mass lost during that phase is
much smaller. We want to mention here at least one
star of which we know that it performed a red-blue evolution and
for which the effect described here almost certainly played a role:
the progenitor of Supernova~1987A (cf. Arnett et al. 1989, Langer 
et al. 1989; cf. also Woosley et al. 1997), 
and the possible twin of it as described by
Brandner et al. (1997).

\section{R4 and the binary scenario}

In this Section, we want to investigate a binary scenario as
possible explanation of the B[e]-phenomenon, which has emerged
while trying to understand the properties of the B[e]~supergiant
R4, on which Franz-Josef Zickgraf has drawn our attention, 
and which is investigated in detail by Zickgraf et al. (1996b).
These authors find that R4 is a spectroscopic binary with a period
of 21.3$\,$yr and an orbital separation of 23 A.U. ($\sim 5000\rso$). 
Presently, the system consists of a B[e] star with a luminosity
of $L=10^5\lso$ and an evolved ($\teff\simeq 9500\,$K) A~star with
$L=1.4\, 10^4\lso$.

These parameters lead immediately to an apparent contradiction:
since the A~star has already evolved off the main sequence,
the almost 10 times more luminous companion should have become
a supernova long time ago. Also, the orbital separation is too large
in order to solve this contradiction by assuming that the A~star
was once more massive and shed mass onto the B~component.

However, another binary scenario may be able to not only explain
the properties of both stars but also the existence of a disk around
the B~star. In this scenario, the system was initially a triple system
consisting of a close pair of stars with an initial mass of $\sim 10\mso$
for each of them, and another $\sim 10\mso$ star surrounding the pair
on a wide orbit. The latter star is identical to the present
A~star, and only serves as a clock without ever interacting with the
close pair.

%=============== Figure binary========================================
\begin{figure}
\centering
\epsfxsize=0.8\hsize
\centerline{\epsffile{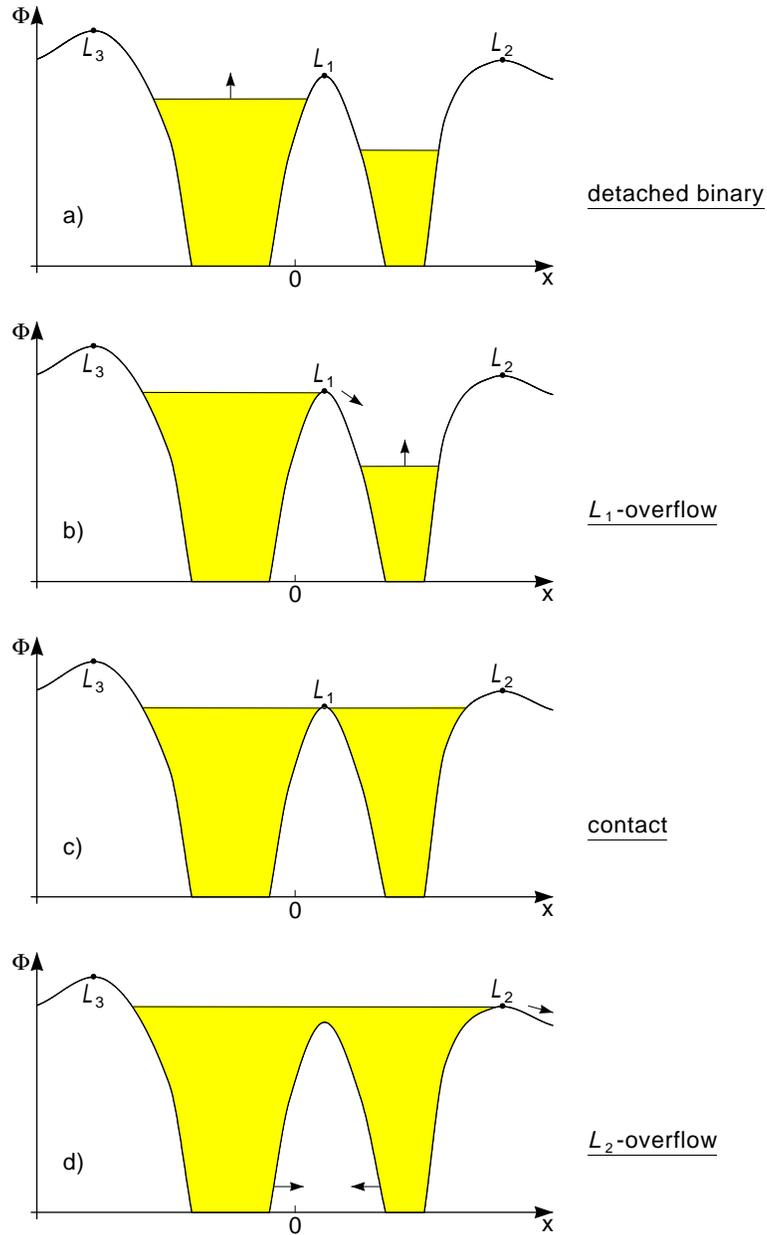}}
  \caption[]{Schematic evolution of the spatial extents of two
   stars of a close binary system which evolves through a contact
   phase and develops mass outflow through the second Lagrange-point
   (``$L_2$-overflow''), which can be responsible for the formation
   of a circumsystem disk or ring (see text for details).}
\end{figure}
%===================================================================

The evolution of the close pair is sketched in Fig.~8. In order to 
end the evolution with the merging of both stars, the more massive
component --- the primary --- must lose mass to the secondary
at a high rate by Roche-lobe overflow through the first 
Lagrange-point $L_{\rm 1}$. If the accretion time scale
$\tau_{\rm \dot M} = M/\dot M$ of the secondary is shorter
than its thermal time scale $\tau_{\rm KH} = GM^2/(R L)$ it will
swell and start filling its own Roche-lobe. As the expansion will
not stop here, matter will soon leave the system through the second 
Lagrange-point $L_{\rm 2}$, and the corresponding angular momentum
loss will bring both stars continuously closer, with the merging
of both stars as final result. Since the outflowing matter leaves
the system in the orbital plane, i.e. through the $L_{\rm 2}$-point
which rotates around the center of mass, it may form a ring or disk-like
structure. 

The most favorable initial conditions for such a scenario may be that
the mass transfer process starts after the primary exhausted
hydrogen in its center, but shortly before its subsequent expansion
leads to the formation of a convective envelope (i.e., late Case~B).
Would the mass transfer start much earlier, i.e. during the core 
hydrogen burning phase of the primary (Case~A), the merger star,
which is now much more massive than the distant companion,
would evolve into a supernova before the latter can leave the main
sequence. Would the mass transfer occur when the primary has a
red supergiant structure (Case~C), both stars might merge as well,
but so quickly that a common envelope is formed which, if at all,
might tend to leave the star rather in a spherically symmetric way.

The merger remnant of a Case~B binary would in fact,
due to its small helium core mass, evolve into a
hot blue supergiant star close to the main sequence in the
HR diagram (cf. Podsiadlowski et al. 1992, Braun \& Langer 1995),
which fits to the effective temperature of the B[e] component
of R4 of $27\, 000\,$K. Due to a strong helium overabundance
in its envelope, it would appear overluminous for its mass,
which seems actually to be observed: Zickgraf et al. (1996b)
find a present mass of $\sim 13\mso$, but derive an initial
mass of  $\sim 20\mso$ from the comparison of its luminosity with
standard single star evolutionary tracks.

\section{Conclusions}

In the previous Sections, we have presented three different
evolutionary scenarios for the formation of a disk-like
structure around a hot evolved massive star. The properties
of the corresponding B[e] candidate models are compared 
in Table~1.

\begin{table}[h]
\begin{center}
\caption{Comparison of observable properties predicted by the three
         B[e] evolutionary scenarios discussed in this paper.}
\begin{tabular}{cccc}
\hline
~~ & very massive main  & supergiant on       & single star         \\
~~ &  sequence star at  & blueward excursion  & remnant of          \\ 
~~ & the $\Omega$-limit & from Hayashi line   & binary merger       \\
\hline
  $\teff$   & ok for most    & ok for less    & scatter in $\teff$  \\
   ~        & luminous B[e]s & luminous B[e]s & plausible           \\
   ~        &      ~         &       ~        &         ~           \\
luminosity  & ok for most    & ok for less    & scatter in $L$      \\
   ~        & luminous B[e]s & luminous B[e]s & plausible           \\
   ~        &      ~         &       ~        &         ~           \\
time scale  & some 10$^5$yr  & some 10$^4$yr  &      (?)            \\
   ~        &      ~         &       ~        &         ~           \\
time integr.&  $\sim 5\mso$  & $\sim 0.1\mso$ &  $\sim 5\mso$(?)    \\
disk mass   &      ~         &       ~        &         ~           \\
\hline
\end{tabular}
\end{center}
\end{table}

First, we can compare the positions of our B[e] candidates
with the observed distribution of
Magellanic Cloud B[e] supergiants in the HR diagram (Gummersbach et
al. 1995; cf. also de Winter \& van den Ancker 1997, and
Zickgraf, this volume).  Clearly, our main
sequence scenario (Sect.~2) would not work for the stars 
with luminosities below $\sim 10^5\lso$. However,
those stars are found to be comparatively cool ($\log
\teff \simeq 4.1$) so that they fit quite well to the 
blue loop scenario outlined in Sect.~3. The very luminous
group of B[e] stars investigated by Gummersbach et al., 
which can not correspond to stars on a blue loop,
are mostly rather hot ($\log \teff \simeq 4.4$) and might
well be main sequence stars. I.e., both scenarios together
cover roughly the range of observed B[e]~supergiants in
the HR diagram. 

For the binary scenario (Sect.~4), there are no quantitative
models, but it appears likely that, assuming a scatter in the
initial orbital parameters and in the initial mass ratio,
the observed scatter in the HR diagram could also be reproduced.

As indicated in Table~1, the time scale of the B[e]~phenomenon
and the time integrated equatorial mass loss are rather different
for the three models. While the main sequence model predicts
a long B[e] life time with several solar masses expelled 
through the disk wind, the blue loop scenario predicts a shorter
life time and less equatorial mass loss.
The expected number of B[e] stars from both scenarios may still
be comparable due to the steep decline of the initial mass function
for larger masses.

B[e] time scale and disk mass loss are least clear for the binary
scenario. During the $L_{\rm 2}$-overflow phase (cf. Fig.~8)
certainly of the order of several solar masses of matter are
lost. However, it is unclear which fraction of that is pushed to
infinity and which fraction forms a disk. Another possibility
which might be considered is that the star which is formed by 
the merger process is an extremely rapid rotator, and consequently
has an equatorially focussed wind. In this case, the time scale
of the B[e] phenomenon would be of the order of the spin-down
time scale of the star.

Finally, we may also consider to combine the three scenarios 
discussed above. For example, if the merger of two stars is
a red supergiant, e.g. in a Case~C system, it may have considerably
more angular momentum than red supergiant formed from single stars.
If such a star evolves into a blue supergiant, our blue loop scenario
must operate again, with an even stronger effect than
in the single star case worked out in Section~3. It may be interesting
to note in this context that Podsiadlowski (1997) suggested 
such a merger scenario for the presupernova evolution of SN~1987A.

\acknowledgements{The authors are very grateful to Peter Conti, Guillermo
Garc\'{\i}a-Segura, Andr\'e Maeder,
Stan Owocki, Philipp Podsiadlowski,
Regina Schulte-Ladbeck, and Franz-Josef Zickgraf 
for enlightening discussions, and to Lars Koesterke
for major help in the production of figures of this paper.
This work has been supported by the Deutsche
Forschungsgemeinschaft through grant No. La~587/15-1}

\end{document}